\newcommand{\be}{\begin{equation}}
\newcommand{\ee}{\end{equation}}
\newcommand{\bea}{\begin{eqnarray}}
\newcommand{\eea}{\end{eqnarray}}
\newcommand{\bq}{\begin{quote}}
\newcommand{\eq}{\end{quote}}

\newcommand{\vecvar}[1]{\mbox{\boldmath$#1$}}

\newcommand{\ri}{{\rm i}}
\newcommand{\rd}{{\rm d}}
\newcommand{\re}{{\rm e}}
\newcommand{\ket}[1]{\left|#1\right\rangle}
\newcommand{\bra}[1]{\left\langle#1\right|}
\newcommand{\braket}[2]{\left\langle#1|#2\right\rangle}
\newcommand{\degree}{\kern-.2em\r{}\kern-.3em}
\newcommand{\degC}{\kern-.2em\r{}\kern-.3em C}
\newcommand{\up}{\uparrow}
\newcommand{\down}{\downarrow}

\newcommand{\eref}[1]{Eq.~(\ref{#1})}

\newcommand{\fref}[1]{Fig.~\ref{#1}}
\newcommand{\Fref}[1]{Figure \ref{#1}}

\documentclass[aps,prb,preprint,superscriptaddress]{revtex4}

\usepackage{graphicx}
\usepackage{float}

\begin{document}

%Title of paper
\title{The ESR intensity and the Dzyaloshinsky-Moriya interaction of 
the nanoscale molecular magnet ${\rm V}_{15}$}

\author{Manabu Machida}
\email[M. Machida: ]{mmachida@umich.edu}
%\homepage[]{Your web page}
%\thanks{}
%\altaffiliation{}
\affiliation{Department of Mathematics, University of Michigan, 
530 Church Street, Ann Arbor, MI 48109, USA}
\author{Toshiaki Iitaka}
\email[T. Iitaka: ]{tiitaka@riken.jp}
\affiliation{Computational Astrophysics Laboratory, RIKEN 
(The Institute of Physical and Chemical Research), 2-1 Hirosawa, Wako, 
Saitama 351-0198, Japan}
\author{Seiji Miyashita}
\email[S. Miyashita: ]{miyashita@phys.s.u-tokyo.ac.jp}
\affiliation{Department of Physics, Graduate School of Science, 
The University of Tokyo, 7-3-1 Hongo, Bunkyo-ku, Tokyo 113-0033, Japan}

\date{\today}

\begin{abstract}
The intensity of electron spin resonance (ESR) of the nanoscale molecular 
magnet ${\rm V}_{15}$ is studied.  We calculate the temperature dependence 
of the intensity at temperatures from high to low.  In particular, we find 
that the low-temperature ESR intensity is significantly affected by the 
Dzyaloshinsky-Moriya interaction.
\end{abstract}

% insert suggested PACS numbers in braces on next line
% insert suggested keywords - APS authors don't need to do this
%\keywords{}

\maketitle

\section{Introduction\label{intro}}

The ${\rm V}_{15}$ molecule has been one of promising nanometer-scale 
molecular magnets since it was first synthesized.
\cite{Muller88a,Muller91a,Gatteschi91a,Barra92a,Gatteschi93a}  
It is the complex of formula 
${\rm K}_6\left[{\rm V}_{15}^{\rm IV}{\rm As}_6{\rm O}_{42}
\left({\rm H}_2{\rm O}\right)\right]\cdot 8{\rm H}_2{\rm O}$.  
In ${\rm V}_{15}$, fifteen vanadium ions of spin $1/2$ form almost 
a sphere.  Three spins in the middle are sandwiched by the upper and 
lower hexagons.  

Different experiments on the magnetization process have shown that 
the magnetization changed adiabatically in a fast sweeping field, and 
a magnetic plateau appeared in a slow sweeping field due to thermal 
bath attached to the molecule.
\cite{Chiorescu00a,Chiorescu00b,Chiorescu03a}  
The latter phenomenon, which is called the phonon bottleneck effect, 
is theoretically analyzed from a general point of view of 
the magnetic Foehn effect.\cite{Saito01a}  
%%%
This smooth change of the magnetization at $H=0{\rm T}$ implies the 
existence of an avoided-level-crossing energy structure. The structure of 
avoided level crossing has been studied. In a model of the triangle Heisenberg 
antiferromagnet with three spins, at $H=0$, two sets of $S=1/2$ doublets 
overlap, and $4$ states degenerate.  
The degeneracy is resolved into two sets of Kramers doublets 
by perturbation such as anisotropy, Dzyaloshinsky-Moriya interaction 
(DMI),\cite{Dzyaloshinsky58a,Moriya60a,Moriya60b,Miyashita01a} 
and also the hyperfine interaction.\cite{Miyashita03}  
Depending on the type of perturbation, there appear many kinds of 
energy structure. 
At the crossing point of the two states of $m=1/2$ of the doublet states 
and one of the quartet state ($m=3/2$), $H=H_{\rm c}$, a kind of avoided 
level crossing is formed. It has been pointed out that adiabatic change 
causes the change of magnetization from $m-1/2$ to $m=1$ because one state 
adiabatically changes to the state with $m=3/2$.\cite{Choi06,Miyashita09}  
In ${\rm V}_{15}$, the equilibrium magnetization curve shows smooth change 
at zero field from $−1/2$ to $1/2$ and at $2.8{\rm T}\,(\equiv H_{\rm c})$ 
from $1/2$ to $3/2$.  ${\rm V}_{15}$ can be described by a triangle model but 
details of the DMI in ${\rm V}_{15}$ are not yet fully understood.

In this paper, first we numerically calculate the temperature 
dependence of the ESR intensity of ${\rm V}_{15}$ using a new 
numerical method (the double Chebyshev polynomial method) of 
calculating the Kubo formula.  We find that the model Hamiltonian for 
${\rm V}_{15}$ including the DMI successfully reproduces the 
experimental temperature dependence of the ESR intensity.  
Second we investigate the ESR at very low temperatures.  
We find that the intensity ratio (the intensity of ${\rm V}_{15}$ divided 
by that of a spin $1/2$) is affected by the DMI at small fields.  
We propose that experimental observation of the intensity ratio enables us 
to estimate the DMI in ${\rm V}_{15}$.  
Finally, we analyze the ESR at low temperatures using a triangle model 
whose energy levels model the low-lying levels of ${\rm V}_{15}$.

\section{Model and Formulation}

\begin{figure}[H]
\begin{center}
\includegraphics[scale=1.5,clip]{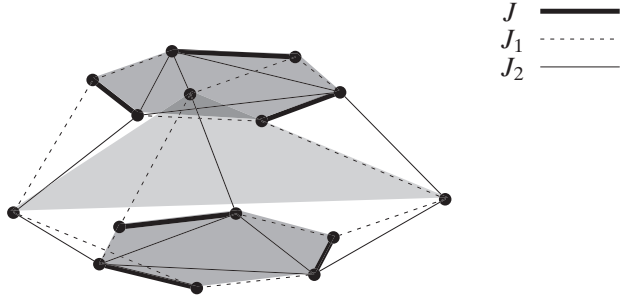}
\end{center}
\caption{Schematic picture of ${\rm V}_{15}$.}
\label{v15schem}
\end{figure}

\Fref{v15schem} shows the structure of vanadium ions in ${\rm V}_{15}$.  
We consider the following spin Hamiltonian for ${\rm V}_{15}$.
\cite{DeRaedt04a,DeRaedt04b,Machida05a}  
\be
\mathcal{H} = 
- \sum_{\langle i,j\rangle}J_{ij}\vecvar{S}_i\cdot\vecvar{S}_j 
+ \sum_{\langle i,j\rangle}\vecvar{D}_{ij}\cdot
\left[\vecvar{S}_i\times\vecvar{S}_j\right]
- \sum_i\vecvar{H}\cdot\vecvar{S}_i.
\label{hami}
\ee
The first term on the right-hand side of \eref{hami} describes the Heisenberg 
interaction.  Coefficients $J_{ij}$ take three values $J$, $J_1$, and 
$J_2$ ($|J|>|J_2|>|J_1|$) depending on the bonds on the upper and lower 
hexagons.  Three spins between two hexagons interact with the hexagons by 
$J_1$ and $J_2$.  There is no interaction among these three spins 
%%% Revised %%%
\cite{Gatteschi91a}.  
%%%%%%%%%%%%%%%
We set $J=-800{\rm K}$, $J_2=-350{\rm K}$, and $J_1=-225{\rm K}$.
\cite{Konstantinidis02a}  The second term describes the DMI.  
We assume the existence of DM vectors $\{\vecvar{D}_{ij}\}$ at the bonds of 
$J$.  In the third term, $\vecvar{H}$ denotes 
the static magnetic field applied to the molecule.  
%%% Revised %%%
We will ignore other effects such as dipolar fields, hyperfine interactions, 
and the crystal field, which are considered to be negligibly small.  
Indeed, the dipolar and hyperfine fields are estimated as $1\,{\rm mK}$ and 
$50\,{\rm mK}$, respectively \cite{Chiorescu00b}.
%%%%%%%%%%%%%%%

\begin{figure}[H]
\begin{center}
\includegraphics[scale=0.7,clip]{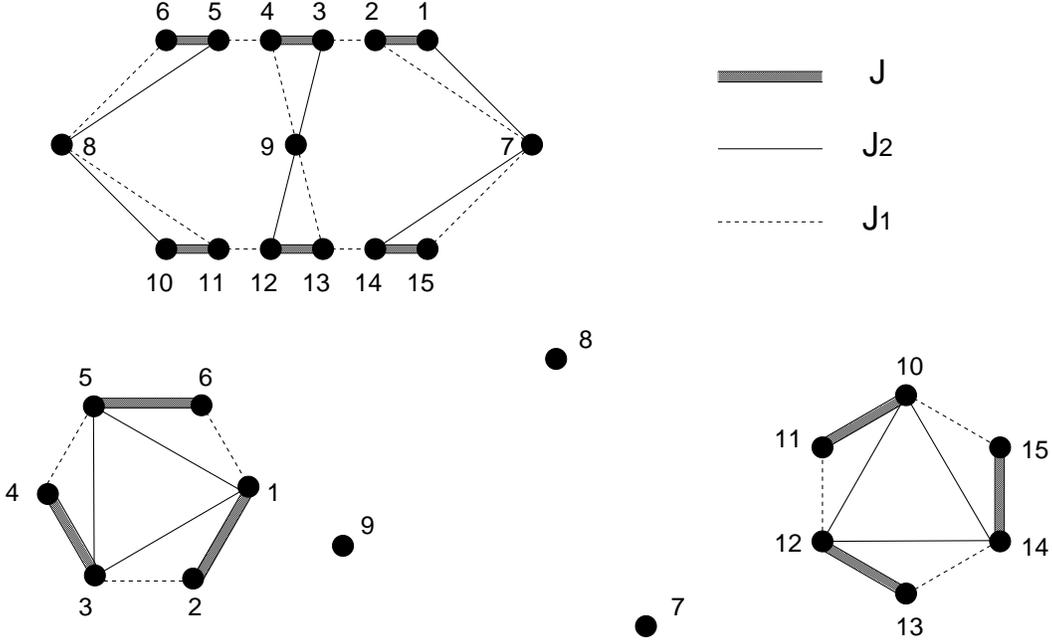}
\end{center}
\caption{Heisenberg interactions between the spins.}
\label{v15bonds}
\end{figure}

\begin{figure}[H]
\begin{center}
\includegraphics[scale=0.7,clip]{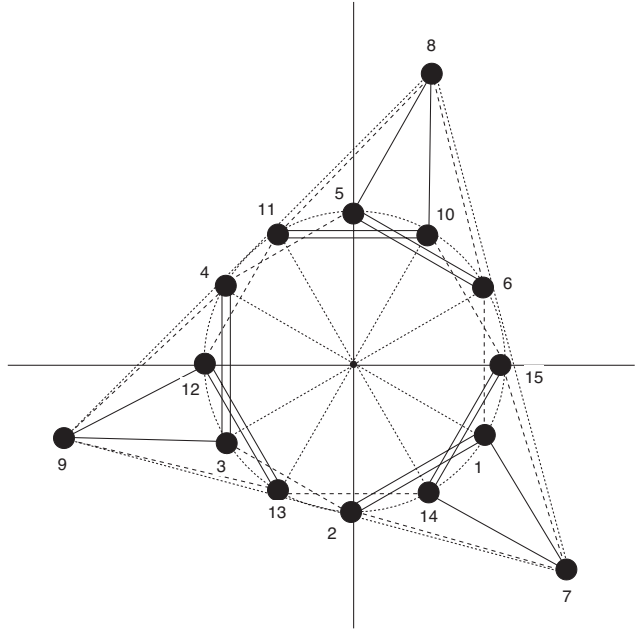}
\end{center}
\caption{The $D_3$ symmetry of ${\rm V}_{15}$.}
\label{v15bonds_d3}
\end{figure}

Figure \ref{v15bonds} explains the interactions between spins.  
If we assume the $D_3$ symmetry of ${\rm V}_{15}$, 
the lower hexagon differs from the upper hexagon by rotation $\pi/6$ 
and we have only one free DM vector, say $\vecvar{D}_{1,2}$ 
(\fref{v15bonds_d3}).  We take the vector $\vecvar{D}_{1,2}$ to be 
$D_{1,2}^x=D_{1,2}^y=D_{1,2}^z=40{\rm K}$.  
%%% Revised %%%
Let us define
\be
\hat{R}(\theta)=
\left(\begin{array}{ccc}
\cos\theta&-\sin\theta&0\\
\sin\theta&\cos\theta&0\\
0&0&1
\end{array}\right),
\qquad
\hat{P}=
\left(\begin{array}{ccc}-1&0&0\\0&1&0\\0&0&-1\end{array}\right).
\ee
We obtain the other DM vectors on the upper hexagon by rotating 
$\vecvar{D}_{1,2}$ by $2\pi/3$ and $4\pi/3$:  
\be
\vecvar{D}_{3,4}=\hat{R}\left(\frac{4\pi}{3}\right)\vecvar{D}_{1,2},\quad
\vecvar{D}_{5,6}=\hat{R}\left(\frac{2\pi}{3}\right)\vecvar{D}_{1,2},
\ee
i.e., $D_{3,4}^x=14.641{\rm K}$, $D_{3,4}^y=-54.641{\rm K}$, 
$D_{3,4}^z=40{\rm K}$, $D_{5,6}^x=-54.641{\rm K}$, $D_{5,6}^y=14.641{\rm K}$, 
and $D_{5,6}^z=40{\rm K}$.  
The DM vectors on the lower hexagon are obtained as 
\be
\vecvar{D}_{10,11}=\hat{P}\vecvar{D}_{1,2},\quad
\vecvar{D}_{12,13}=\hat{R}\left(\frac{2\pi}{3}\right)\vecvar{D}_{10,11},\quad
\vecvar{D}_{14,15}=\hat{R}\left(\frac{4\pi}{3}\right)\vecvar{D}_{10,11},
\ee
i.e.,
$D_{10,11}^x=-40{\rm K}$, $D_{10,11}^y=40{\rm K}$, $D_{10,11}^z=-40{\rm K}$, 
$D_{12,13}^x=-14.641{\rm K}$, $D_{12,13}^y=-54.641{\rm K}$, 
$D_{12,13}^z=-40{\rm K}$, 
$D_{14,15}^x=54.641{\rm K}$, $D_{14,15}^y=14.641{\rm K}$, and 
$D_{14,15}^z=-40{\rm K}$.  
%%%%%%%%%%%%%%%

Let us calculate the ESR intensity of ${\rm V}_{15}$ using the Hamiltonian 
(\ref{hami}).  By the Kubo formula\cite{Kubo54a,Kubo57a}, the imaginary part 
of the susceptibility $\chi''(\omega,T)$ on the transverse field 
$\vecvar{H}$ parallel to the $x$-axis is given by the Fourier transform of 
the spin-spin correlation function; 
\be
\chi''(\omega;T) = \left(1-\re^{-\beta\omega}\right){\rm Re}
\int_0^{\infty} \langle M^xM^x(t)\rangle \re^{-\ri\omega t} \rd t,
\ee
where $\omega$ is the frequency of the radiation field, $T$ is temperature, 
and $\beta=1/T$.  The absorption is given by 
\be
I(\omega;T)=\frac{\omega H_{\rm R}^2}{2}\chi''(\omega;T),
\ee
where $H_{\rm R}$ is the amplitude of the radiation field.  
We obtain the ESR intensity $I(T)$ by integrating $I(\omega,T)$ 
with respect to $\omega$; 
\be
I(T)=\int_0^{\infty}I(\omega,T) \rd\omega.
\ee

In the present paper, we obtain $I(T)$ without diagonalization.  
Our method can readily be applied to other nanomagnets.  
In Appendix \ref{DCEM}, we describe the numerical method of 
calculating the Kubo formula for huge Hamiltonian quantum many-body 
systems.\cite{Machida05a}  This method, which we call the double Chebyshev 
expansion method (DCEM), realizes $O(N)$ calculation in both speed and 
memory.  In the method, we evaluate the summation over all the states in the 
expression of the correlation function by using the average over 
a few number of random initial states.  Furthermore, we calculate the 
exponential operators $\re^{-\beta\mathcal{H}}$ and $\re^{-\ri\mathcal{H}t}$ 
by expanding them in the Chebyshev polynomial.  
The DCEM takes all the states in the Hilbert space into account 
and enables us to obtain the ESR at any temperature.  

\section{Temperature Dependence of the ESR Intensity\label{Ajiro}}

\begin{figure}[H]
\begin{center}
\includegraphics[scale=1.5,clip]{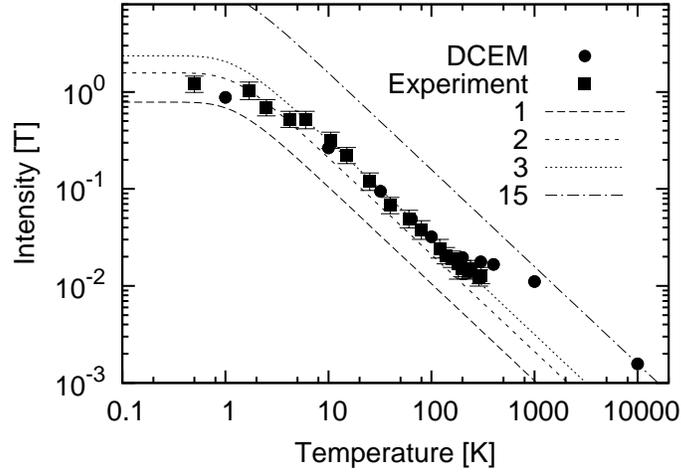}
\end{center}
\caption{The temperature dependences of the intensity for 
$\vecvar{H}||z$ and $H=2{\rm T}$ and the experimental intensity.}
\label{ajiro-2T-DM}
\end{figure}

We study the temperature dependence of the ESR intensity of 
${\rm V}_{15}$.  We apply the magnetic field parallel to the $c$-axis 
of the molecule ($z$-axis): $\vecvar{H}=(0,0,H)$.  
\Fref{ajiro-2T-DM} shows intensities at $H=2{\rm T}$ from $1{\rm K}$ to 
$10000{\rm K}$. The intensity obtained by the DCEM (solid circles) is plotted 
together with experimental data by Ajiro, et al.\cite{Ajiro03a} 
(solid squares).  The intensity by the DCEM in \fref{ajiro-2T-DM} agrees 
with the experimental data.  In addition, the dashed line denotes the 
intensity 
\be
I_1(T)=H\tanh{\frac{\beta H}{2}}
\label{I1}
\ee
of an isolated spin $1/2$.  
The short-dashed line, dotted line, and dash-dotted line denote 
$2I_1(T)$, $3I_1(T)$, and $15I_1(T)$ for non-interacting 2, 3, and 15 
spins, respectively.  

When the temperature is sufficiently higher than the interactions, 
spins in ${\rm V}_{15}$ are almost isolated.  Therefore, the intensity is 
expected to meet the dash-dotted line at very high temperatures.  
In \fref{ajiro-2T-DM}, the intensity starts to deviate from the dotted line 
near $200{\rm K}$.  As the temperature decreases, the effective number of 
spins changes from 15 to 3, and the intensity stays on the dotted line 
at temperatures between $10{\rm K}$ and $100{\rm K}$.  

\begin{figure}[H]
\begin{center}
\includegraphics[scale=1.5,clip]{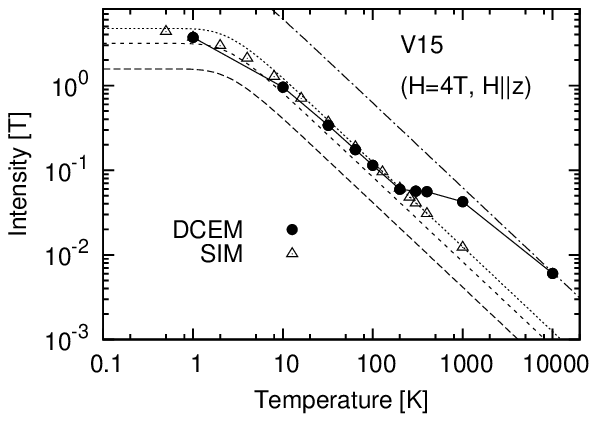}
\end{center}
\caption{The temperature dependence of the intensity for 
$\vecvar{H}||z$ and $H=4{\rm T}$.}
\label{ajiro-4T-DM}
\end{figure}

\Fref{ajiro-4T-DM} is the same as \fref{ajiro-2T-DM} except $H=4{\rm T}$.  
At temperatures above $10{\rm K}$, the intensity in 
\fref{ajiro-4T-DM} behaves almost the same as that in 
\fref{ajiro-2T-DM}.  
However, the temperature dependence of the effective number is 
different from that in \fref{ajiro-2T-DM} below $10{\rm K}$.  
In \fref{ajiro-4T-DM}, the effective number of spins changes from 
15 to 3 at high temperatures, and stays on the dotted line for $3I_1(T)$ as 
temperature decreases.  A similar behavior of the low-temperature intensity 
has also been predicted.\cite{Sakon04a}

\section{The Low-Temperature Intensity Ratio}

\subsection{$I(T)$ at low temperatures}

\begin{figure}[H]
\begin{center}
\includegraphics[scale=1.5,clip]{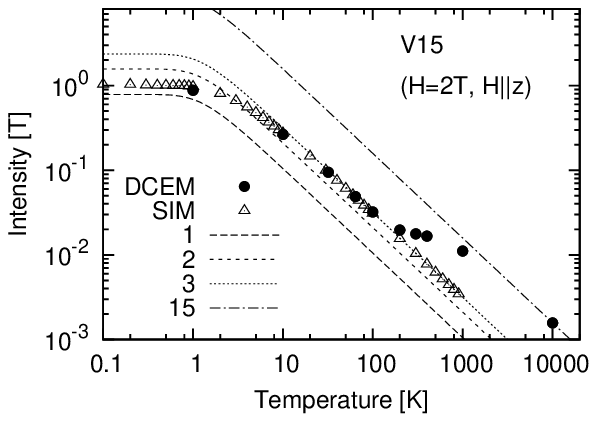}
\end{center}
\caption{The temperature dependence of the intensity for 
$\vecvar{H}||z$ and $H=2{\rm T}$.}
\label{sim-dcem}
\end{figure}

Although the DMI affects $I(T)$ only mildly at $T>1{\rm K}$, it significantly 
changes $I(T)$ at low temperatures.  At low temperatures, only transitions 
among low-lying energy levels near the ground state are responsible for 
the ESR absorption.  This fact enables us to obtain the intensity by 
investigating the transitions among the lowest eight levels.  
Thus, we can calculate intensity with the subspace iteration method 
(SIM).\cite{Machida05a,Chatelin88a,Mitsutake96a}  The response 
$\chi''(\omega,T)$ is obtained by direct diagonalization in the small 
reduced space.  In \fref{sim-dcem}, we compare the result of 
\fref{ajiro-2T-DM} and the data obtained by the SIM.  We find a good 
agreement below $100{\rm K}$.  Above this temperature, the lowest eight 
states are not enough to represent the system.  

\begin{figure}[H]
\begin{center}
\includegraphics[scale=1.2,clip]{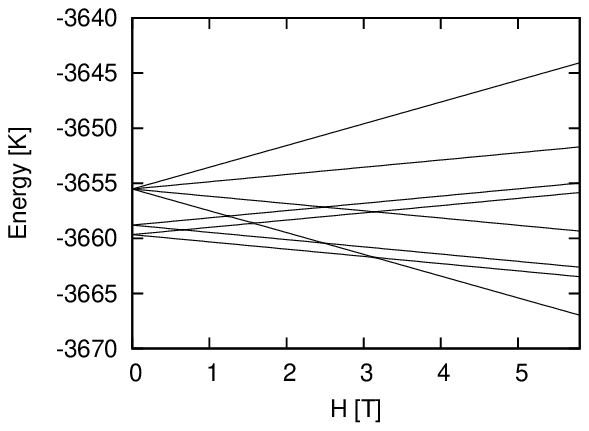}
\includegraphics[scale=1.2,clip]{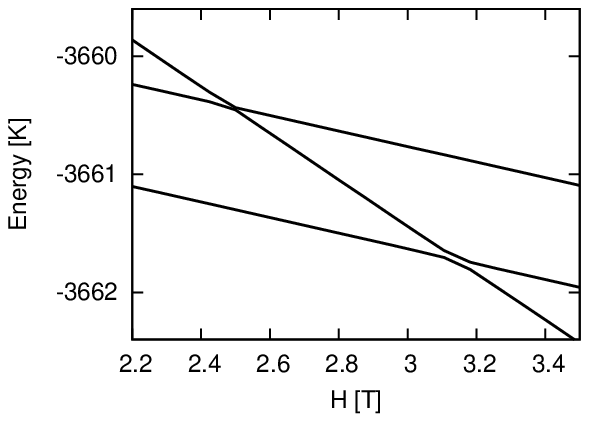}
\end{center}
\caption{(a) The lowest eight levels of ${\rm V}_{15}$ as a function of 
$\vecvar{H}||z$.  (b) Magnified figure near $3{\rm T}$.}
\label{lev-Hsz}
\end{figure}

In \fref{lev-Hsz}(a), we show the lowest eight energy levels of ${\rm V}_{15}$ 
as a function of the magnetic field $H$ applied in the 
$z$-direction.  Due to the DMI, we have an avoided level crossing near 
$3{\rm T}$ as shown in \fref{lev-Hsz}(b).  Without the DMI, the gap closes and 
the two levels just cross each other.  

\begin{figure}[H]
\begin{center}
\includegraphics[scale=1.5,clip]{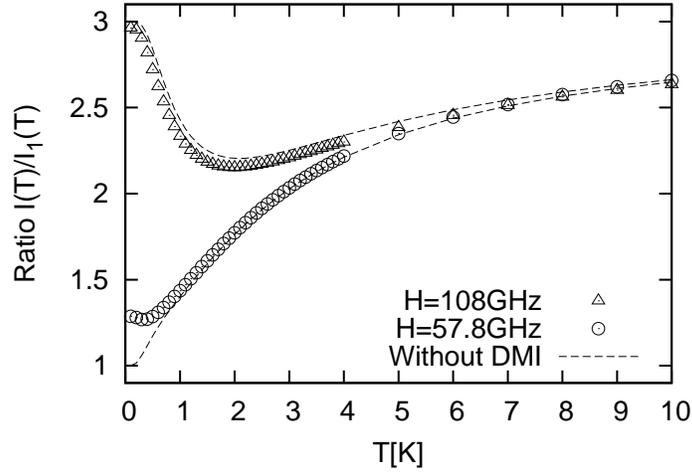}
\end{center}
\caption{The intensity ratios of ${\rm V}_{15}$ as a function of the 
temperature.  The intensity ratios without the DMI are also shown for 
comparison.}
\label{ratio-v15}
\end{figure}

We define the intensity ratio $R(T)$ as
\be
R(T)=I(T)/I_1(T),
\ee
where $I_1(T)$ is the intensity of a single spin (\eref{I1}).  
In the previous section, we have found $R(T)\simeq 3$ for $H=4{\rm T}$ and 
$R(T)$ is slightly larger than $1$ for $H=2{\rm T}$ in the low temperature 
limit.  \Fref{ratio-v15} shows $R(T)$ in a low temperature region.  
In \fref{ratio-v15}, the circles and triangles show $R(T)$ at 
$H=57.8{\rm GHz}$ ($\simeq 2{\rm T}$) and at $H=108{\rm GHz}$ 
($\simeq 4{\rm T}$), respectively.  The dashed lines in \fref{ratio-v15} 
show the ratios without the DMI at $H=2{\rm T}$ and at $H=4{\rm T}$ for 
comparison.  
First, we notice that $H_{\rm c}$ can be experimentally estimated with 
the low-temperature ESR by measuring the field at which the destination of 
the intensity changes.  
Second, we see that $R(T=0{\rm K})$ at $H=2{\rm T}$ deviates from $1$ due to 
the DMI, while $R(T=0{\rm K})$ at $H=4{\rm T}$ stays very close to the 
dotted line.  Thus it would be possible to experimentally determine the DMI 
in ${\rm V}_{15}$ by observing the intensity at low temperatures.  

\begin{figure}[H]
\begin{center}
\includegraphics[scale=1.2,clip]{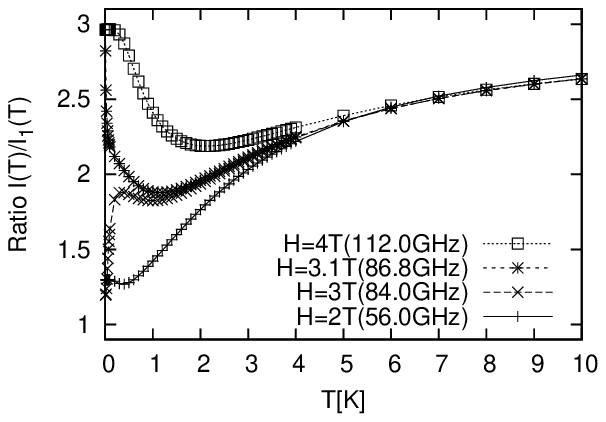}
\includegraphics[scale=1.2,clip]{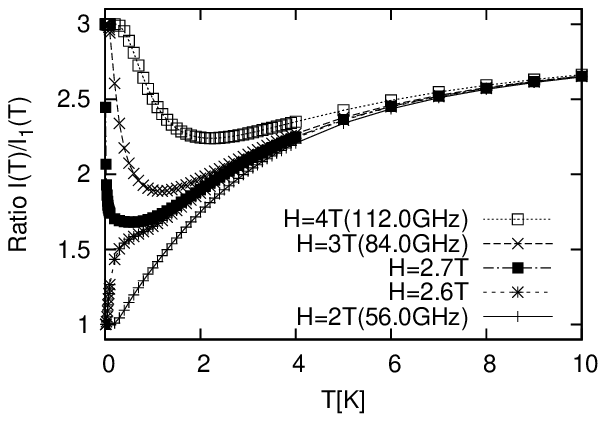}
\end{center}
\caption{(a) Intensity ratios of ${\rm V}_{15}$ as a function of 
temperature for various fields.  (b) Intensity ratios without the DMI.}
\label{ratio-DM}
\end{figure}

Furthermore, we study intensity ratios at various fields near the 
avoided level crossing point (hereafter we refer to this field as 
$H_{\rm c}$) in the presence (\fref{ratio-DM}(a)) and absence 
(\fref{ratio-DM}(b)) of the DMI.  The ratio goes to 3 as $T\to0$ for a field 
higher than $H_{\rm c}$.  In contrast, it goes near 1 for a lower field.  
The derivative of the line for $H=2{\rm T}$ in \fref{ratio-DM}(a) is 
negative at $0{\rm K}$, while that for $H=3{\rm T}$ is positive.  Hence the 
line for $H=2{\rm T}$ has a dip at $T\simeq 0.5{\rm K}$.  In the absence of 
the DMI, the line for $H=2{\rm T}$ does not have a dip.  

\subsection{Triangle Model}

As shown in Figs.~\ref{ratio-v15} and \ref{ratio-DM}(a), the intensity 
ratio for lower fields deviates from 1 in the low-temperature limit.  
Here, we study the mechanism of this deviation analytically in a triangle 
model of three spins which well describes the low-lying energy structure of 
${\rm V}_{15}$.

\begin{figure}[H]
\begin{center}
\includegraphics[scale=1.5,clip]{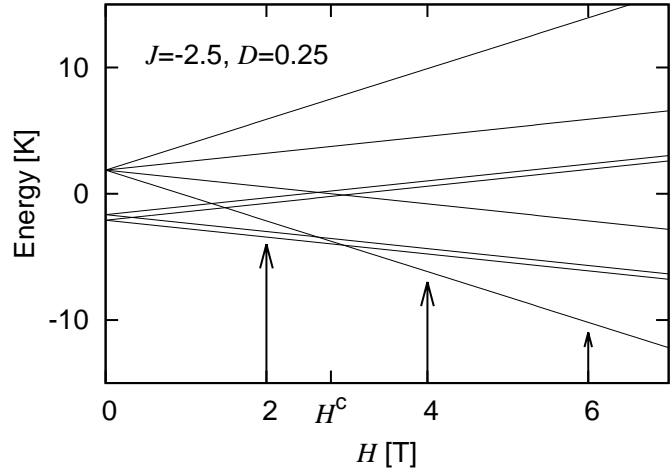}
\end{center}
\caption{Energy levels of the triangle model as a function of the field $H$.}
\label{levtri}
\end{figure}

The triangle model\cite{DeRaedt04b} is given by the Hamiltonian \eref{hami} 
with $J_{ij}=J<0$, $\vecvar{H}=(0,0,H)$, $H>0$, 
\be
\begin{array}{ll}
D_{12}^x=D_x, & D_{12}^y=D_y, \\
D_{23}^x=(-D_x+\sqrt{3}D_y)/2, & D_{23}^y=(-\sqrt{3}D_x-D_y)/2, \\
D_{31}^x=(-D_x-\sqrt{3}D_y)/2, & D_{31}^y=(\sqrt{3}D_x-D_y)/2, 
\end{array}
\label{tridmxy}
\ee
and 
\be
D_{12}^z=D_{23}^z=D_{31}^z=D_z.
\label{tridmz}
\ee
Note that the elements of the DM vectors are chosen so that the 
model satisfies the $C_3$ symmetry.  
If we put $J=-2.5{\rm K}$ and $D_x=D_y=D_z=0.25{\rm K} (\equiv D)$, 
energy levels of the triangular model well reproduce the lowest 
eight levels of ${\rm V}_{15}$ as shown in \fref{levtri}.  
In the absence of the DMI, the ground state 
magnetization $M^z$ changes from $1/2$ to $3/2$ at 
\be
H_{\rm c} \equiv \frac{3}{2}|J| \simeq 2.8{\rm T}.
\ee

The intensity ratio $R_3(T)$ of this triangle model is obtained as 
\begin{widetext}
\bea
R_3(T) &=& \left[I_1(T)\sum_m\re^{-\beta E_m}\right]^{-1}
\sum_{m} r(m) \nonumber \\
r(m) &=& \sum_{n\,(E_n>E_m)}(E_n-E_m)
\left(\re^{-\beta E_m}-\re^{-\beta E_n}\right)
\left|2\bra{w_m}M_x\ket{w_n}\right|^2,
\label{R3def}
\eea
\end{widetext}
where $\{E_m\}$ and $\{\ket{w_m}\}$ are the eigenvalues and eigenvectors 
of the triangle-model Hamiltonian, respectively.  
In Appendix \ref{detailtri}, $R_3(T)$ at low temperatures around $H_{\rm c}$ 
is calculated.

Let us explore the nonmonotonic temperature dependence of the intensities 
shown in \fref{ratio-DM} by using the triangle model with different values 
of $D_x$, $D_y$, and $D_z$.  

First, we consider how the DMI causes 
the dip at $T\simeq 0.5{\rm K}$ for $H=4{\rm T}$ in \fref{ratio-DM}(a).  
In the absence of the DMI (\fref{ratio-DM}(b)), by setting 
$D_x=D_y=D_z=0$, we obtain $\rd R_3(T)/\rd T>0$ for $H<H_{\rm c}$ and 
$\rd R_3(T)/\rd T<0$ for $H_{\rm c}<H$ 
(see \eref{R3derivnoDM} in Appendix \ref{detailtri}).  
To investigate the intensities in \fref{ratio-DM}(a), we set 
$D_x=D_y=0,\,D_z\neq 0$ (uniaxial).  Note that 
the structure of the energy levels with $D_x=D_y=D_z\neq 0$ is the 
same as that in the uniaxial case only except that the gap closes in 
the uniaxial case.  We find that $\rd R_3(T)/\rd T<0$ for 
$H\lesssim H_{\rm c}$  and $\rd R_3(T)/\rd T>0$ for $H\simeq H_{\rm c}$ 
(see \eref{R3deriv} in Appendix \ref{detailtri}), which implies a dip at 
a low field.  

Next, let us consider the effect of a weak DMI.  
We set $D_x=D_y=D_z=D\neq 0$.  
By ignoring smaller terms than $O(D)$, we have (Appendix \ref{detailtri})
\be
R_3(T)\simeq\left\{\begin{array}{ll}
1+\frac{\Delta}{H}-\frac{2\Delta}{H}\,\re^{-\Delta/T} 
& (H<H_{\rm c}) \\
3-\left(2-\frac{\Delta}{H}\right)\,
\re^{-|H-H_{\rm c}-\Delta/2|/T} 
& (H_{\rm c}<H),
\end{array}\right.
\label{weakR3}
\ee
where we defined 
\be
\Delta\equiv\sqrt{3}D.
\ee
Taking the limit $T\to 0$, we have 
\be
R_3(T)\simeq\left\{\begin{array}{ll}
1+\frac{\Delta}{H} & (H<H_{\rm c}) \\
3                          & (H_{\rm c}<H).
\end{array}\right.
\ee
This term $\Delta/H$ describes the deviation of $R(T)$ from 1, and 
the deviation is proportional to $D$.

\section{Summary}
In this paper, we studied the ESR of the nanomagnet ${\rm V}_{15}$.  
We investigated the temperature dependence of the intensity on the DMI for 
different values of the static field $H$.  The DMI significantly affects the 
low-temperature intensity of ${\rm V}_{15}$.  We found that the intensity 
at $H<H_{\rm c}$ has a dip as a function of temperature due to the DMI.  
We analyzed the dip using the three-spin model and obtained analytical 
expressions of the intensity.  Experimental measurement of the dip 
may be used to estimate the DMI of ${\rm V}_{15}$.

\begin{acknowledgments}
The authors would like to thank Professor Hans De Raedt for valuable 
discussions. The simulations were partially carried out by using the 
computational facilities of the Super Computer Center of Institute 
for Solid State Physics, The University of Tokyo, and 
the Advanced Center for Computing and Communication, RIKEN 
(The Institute of Physical and Chemical Research).
\end{acknowledgments}

\appendix

\section{DCEM\label{DCEM}}
The DCEM is an extension of the 
Boltzmann-weighted time-dependent method (BWTDM).\cite{Iitaka97a,Iitaka03a}  
The DCEM differs from the BWTDM by the third step below.  
The procedure of the DCEM is divided by the following five steps.  

At the first step, we prepare a random vector 
$\ket{\Phi}$.\cite{DeRaedt00a,Iitaka04a}  
For a given basis $\ket{n}$ of the Hilbert space, this 
random vector is given by 
$\ket{\Phi}=\sum_{n=1}^N\ket{n}\re^{\ri\theta_n}$.  
Here, 
the dimension of the Hilbert space is $N$ and the random numbers 
$\{\theta_n\}$ take values from $-\pi$ to $\pi$.  

At the second step, we obtain the Boltzmann-weighted random vector 
$\ket{\Phi_{\rm Boltz}}=\re^{-\beta\mathcal{H}/2}\ket{\Phi}$ 
by expanding it with the Chebyshev polynomial; 
\bea
\re^{-\beta\mathcal{H}/2} &=& 
I_0\left(-\beta\Delta\lambda/2\right)T_0(\mathcal{H}_{\rm sc}) 
\nonumber \\
&+& 2\sum_{k}
I_k\left(-\beta\Delta\lambda/2\right)T_k(\mathcal{H}_{\rm sc}),
\eea
where $I_k(x)$ is the modified Bessel function and 
$T_k(\mathcal{H}_{\rm sc})$ is the Chebyshev polynomial, which satisfies 
$T_k(\mathcal{H}_{\rm sc})=
2\mathcal{H}_{\rm sc}T_{k-1}(\mathcal{H}_{\rm sc})-
T_{k-2}(\mathcal{H}_{\rm sc})$, 
$T_0(\mathcal{H}_{\rm sc})=1$, and 
$T_1(\mathcal{H}_{\rm sc})=\mathcal{H}_{\rm sc}$.  
We note that the eigenvalues of 
$\mathcal{H}_{\rm sc}\,(=\mathcal{H}/\Delta\lambda)$ 
are confined between $-1$ to $1$.  
By this procedure, we obtain vectors $\re^{-\beta\mathcal{H}/2}\ket{\Phi}$ 
without diagonalization.  

At the third step, we obtain 
$\ket{\Phi_{\rm Boltz};t}=\re^{-\ri\mathcal{H}t}\ket{\Phi_{\rm Boltz}}$ 
and 
$\ket{\Phi_{M^x};t}=\re^{-\ri\mathcal{H}t}\ket{\Phi_{M^x}}$, where 
$\ket{\Phi_{M^x}}=M^x\ket{\Phi_{\rm Boltz}}$ and 
$M^x=\sum S^x_j$.  In the BWTDM, 
the time evolution is performed by the leap frog method, 
which evolves a state $\ket{\phi;t}$ as 
\be
\ket{\phi;t+\Delta t}=
-2\ri\mathcal{H}\Delta t\ket{\phi;t}+\ket{\phi;t-\Delta t}.  
\ee
Note that the condition $E_{\rm max}\Delta t\ll 1$ should be satisfied, 
where $E_{\rm max}$ is the largest eigenvalue of the Hamiltonian.  
In the DCEM, we make use of the Chebyshev polynomial expansion 
in order to obtain the time evolution of the vector; 
\bea
\ket{\phi;t+\tau} &=& 
\re^{-\ri\tau\Delta\lambda\mathcal{H}_{\rm sc}}\ket{\phi;t} \nonumber \\
&=& J_0(\tau\Delta\lambda)T_0(\mathcal{H}_{\rm sc})\ket{\phi;t} 
\nonumber \\
&+& 2\sum_{k} (-\ri)^k
J_k(\tau\Delta\lambda)T_k(\mathcal{H}_{\rm sc})\ket{\phi;t},
\eea
where $J_k(x)$ is the Bessel function.  Note that the time step $\tau$ is 
not necessarily small.  
In the ESR experiment for ${\rm V}_{15}$, the magnetic field $H$ 
$(\sim 1 {\rm K})$ is usually much smaller than the strongest coupling 
$|J|$ $(\sim 10^3 {\rm K})$.  Hence the frequency 
of precession of the spins is small.   This means that 
we need to evolve state vectors for a long time but do not need fine 
resolution of the time step.  This is why the DCEM is more efficient than 
the BWTDM for the ESR of ${\rm V}_{15}$.  

At the fourth step, we calculate the correlation function 
\bea
\langle M^xM^x(t)\rangle 
&=& 
\frac{
{\rm Tr} \re^{-\beta\mathcal{H}}M^x\re^{\ri\mathcal{H}t}M^x
\re^{-\ri\mathcal{H}t}
}{
{\rm Tr} \re^{-\beta\mathcal{H}}
} \nonumber \\
&=& 
\frac{
\left[\bra{\Phi_{M^x};t}M^x\ket{\Phi_{\rm Boltz};t}\right]_{\rm av}
}{
\left[\braket{\Phi_{\rm Boltz}}{\Phi_{\rm Boltz}}\right]_{\rm av}
},
\eea
where the traces are replaced by the ensemble averages 
($\left[\cdot\right]_{\rm av}$) with respect to the random 
vectors; for any operator $\hat{X}$, ${\rm Tr}\hat{X}$ is calculated as 
\bea
\left[\bra{\Phi}\hat{X}\ket{\Phi}\right]_{\rm av}
&=&
\sum_n\bra{n}\hat{X}\ket{n}
+\sum_{m,n}\left[\re^{\ri(\theta_m-\theta_n)}-\delta_{mn}\right]_{\rm av}
\bra{n}\hat{X}\ket{m} \nonumber \\
&\simeq& {\rm Tr}\hat{X}.
\eea

Finally, $\chi''(\omega;T)$ is obtained by the Fourier transform of 
$\langle M^xM^x(t)\rangle$.  
\bea
\chi''(\omega;T) &=& \left(1-\re^{-\beta\omega}\right){\rm Re}
\int_0^{\infty} \langle M^xM^x(t)\rangle \re^{-\ri\omega t} \rd t \nonumber \\
&\hspace{-30mm}=& \hspace{-15mm}\left(1-\re^{-\beta\omega}\right){\rm Re}
\int_0^{T_{\rm max}} \langle M^xM^x(t)\rangle \re^{-\ri\omega t} 
\re^{-\eta^2t^2/2} \rd t.
\eea
Here we introduced the Gaussian filter with variance $1/\eta^2$.  
This $\eta$ determines the frequency resolution.  The upper limit of the 
integral $T_{\rm max}$ satisfies 
$T_{\rm max}\sim 1/\eta$ in order to avoid the Gibbs oscillation.  
Also $\eta$ should satisfy the conditions that 
$0<\eta\ll 1$, $\eta\ll H$, and $\beta\eta^2\ll H$.

\section{Details of the Triangle Model\label{detailtri}}

We obtain the block-diagonalized form of the Hamiltonian 
by using the following basis vectors $\{\ket{v_j}\}$.  
\bea
\ket{v_1}&=& \frac{-1}{2\sqrt{3}}\left[
(1+\ri\sqrt{3})\ket{\up\up\down}+(1-\ri\sqrt{3})\ket{\up\down\up}
-2\ket{\down\up\up}\right], \nonumber \\
\ket{v_2}&=& \frac{-1}{2\sqrt{3}}\left[
(1-\ri\sqrt{3})\ket{\up\up\down}+(1+\ri\sqrt{3})\ket{\up\down\up}
-2\ket{\down\up\up}\right], \nonumber \\
\ket{v_3}&=& \frac{-1}{2\sqrt{3}}\left[
-2\ket{\up\down\down}+(1+\ri\sqrt{3})\ket{\down\up\down}
+(1-\ri\sqrt{3})\ket{\down\down\up}\right], \nonumber \\
\ket{v_4}&=& \frac{-1}{2\sqrt{3}}\left[
-2\ket{\up\down\down}+(1-\ri\sqrt{3})\ket{\down\up\down}
+(1+\ri\sqrt{3})\ket{\down\down\up}\right], \nonumber \\
\ket{v_5}&=& \ket{\up\up\up}, \nonumber \\
\ket{v_6}&=& \frac{1}{\sqrt{3}}\left[
\ket{\up\up\down}+\ket{\up\down\up}+\ket{\down\up\up}\right], \nonumber \\
\ket{v_7}&=& \frac{1}{\sqrt{3}}\left[
\ket{\up\down\down}+\ket{\down\up\down}+\ket{\down\down\up}\right], 
\mbox{ and} \nonumber \\
\ket{v_8}&=& \ket{\down\down\down}.
\label{vi}
\eea
Now we have
\be
\left\{\begin{array}{lll}
\mathcal{H}\ket{v_5} &=& -\frac{1}{4}(3J+6H)\ket{v_5}
+\frac{3}{4}D_+\ket{v_1}, \\
\mathcal{H}\ket{v_1} &=& \frac{3}{4}D_-\ket{v_5}
+\frac{1}{4}(3J-2H-2\sqrt{3}D_z)\ket{v_1},
\end{array}\right. 
\ee
\be
\left\{\begin{array}{lll}
\mathcal{H}\ket{v_4} &=& \frac{1}{4}(3J+2H+2\sqrt{3}D_z)\ket{v_4}
+\frac{\sqrt{3}}{4}D_-\ket{v_6}, \\
\mathcal{H}\ket{v_6} &=& \frac{\sqrt{3}}{4}D_+\ket{v_4}-
\frac{1}{4}(3J+2H)\ket{v_6},
\end{array}\right. 
\ee
\be
\left\{\begin{array}{lll}
\mathcal{H}\ket{v_2} &=& \frac{1}{4}(3J-2H+2\sqrt{3}D_z)\ket{v_2}
-\frac{\sqrt{3}}{4}D_+\ket{v_7}, \\
\mathcal{H}\ket{v_7} &=& -\frac{\sqrt{3}}{4}D_-\ket{v_2}
+\frac{1}{4}(-3J+2H)\ket{v_7},
\end{array}\right.
\ee
\be
\left\{\begin{array}{lll}
\mathcal{H}\ket{v_8} &=& \frac{1}{4}(-3J+6H)\ket{v_8}
+\frac{3}{4}D_-\ket{v_3}, \mbox{ and}\\
\mathcal{H}\ket{v_3} &=& \frac{3}{4}D_+\ket{v_8}
+\frac{1}{4}(3J+2H-2\sqrt{3}D_z)\ket{v_3},
\end{array}\right.
\ee
where $D_{\pm}=D_x\pm\ri D_y$.

\subsubsection{Uniaxial Dzyaloshinsky-Moriya interaction}

In the presence of the uniaxial DMI ($D_x=D_y=0$), 
the triangle-model Hamiltonian is diagonalized by the vectors 
$\{\ket{v_j}\}$, i.e., $\{\ket{v_j}\}$ form the eigenvectors of 
the Hamiltonian.  The correspondent eigenvalues $\{E_j\}$ are given by 
\bea
E_8 &=& -\frac{3}{4}J+\frac{3}{2}H, \nonumber \\
E_7 &=& -\frac{3}{4}J+\frac{1}{2}H, \nonumber \\
E_6 &=& -\frac{3}{4}J-\frac{1}{2}H, \nonumber \\
E_5 &=& -\frac{3}{4}J-\frac{3}{2}H, \nonumber \\
E_4 &=& \frac{3}{4}J+\frac{1}{2}H+\frac{\sqrt{3}}{2}D_z, \nonumber \\
E_3 &=& \frac{3}{4}J+\frac{1}{2}H-\frac{\sqrt{3}}{2}D_z, \nonumber \\
E_2 &=& \frac{3}{4}J-\frac{1}{2}H+\frac{\sqrt{3}}{2}D_z, 
\mbox{ and} \nonumber \\
E_1 &=& \frac{3}{4}J-\frac{1}{2}H-\frac{\sqrt{3}}{2}D_z.
\eea
Besides, nonzero matrix elements are obtained as 
\bea
\bra{v_1}M^x\ket{v_4} &=& -\frac{1}{2}, \nonumber \\
\bra{v_2}M^x\ket{v_3} &=& -\frac{1}{2}, \nonumber \\
\bra{v_5}M^x\ket{v_6} &=& \frac{\sqrt{3}}{2}, \nonumber \\
\bra{v_6}M^x\ket{v_7} &=& 1, \mbox{ and} \nonumber \\
\bra{v_7}M^x\ket{v_8} &=& \frac{\sqrt{3}}{2},
\eea
where 
\be
M^x=S_1^x+S_2^x+S_3^x.
\ee
Note that $M^x$ is hermitian, i.e., 
$\bra{v_i}M^x\ket{v_j}=(\bra{v_j}M^x\ket{v_i})^*$.

The magnitude relation of $E_1$, $E_2$, and $E_5$ depends on 
$H$: 
\be
\left\{\begin{array}{ll}
E_1 < E_2 < E_5 & (H<H_{\rm c}-\frac{\Delta_z}{2}), \\
E_1 < E_5 < E_2 & (H_{\rm c}-\frac{\Delta_z}{2}<H<
H_{\rm c}+\frac{\Delta_z}{2}), \mbox{ and} \\
E_5 < E_1 < E_2 & (H_{\rm c}+\frac{\Delta_z}{2}<H),
\end{array}\right.
\ee
where we define 
\be
\Delta_z\equiv\sqrt{3}D_z.
\ee
In the absence of the DMI ($D_x=D_y=D_z=0$), the two-fold degenerate 
ground state ($\ket{v_1}, \ket{v_2}$) of $M_z=1/2$ and 
the state ($\ket{v_5}$) of $M_z=3/2$ intersect at 
$H=H_{\rm c}$.  
In the uniaxial DMI, the states $\ket{v_2}$ and $\ket{v_5}$ intersect at 
$H=H_{\rm c}-\Delta_z$, and 
the states $\ket{v_1}$ and $\ket{v_5}$ intersect at 
$H=H_{\rm c}+\Delta_z$, respectively.  

Note that, near $H_{\rm c}$, we calculate the intensity ratio by 
taking into account only three low-lying states $\ket{v_1}$, $\ket{v_2}$, 
and $\ket{v_5}$ at very low temperatures, and we have 
\bea
\sum_m\re^{-\beta E_m} &\simeq& 
\re^{-\beta E_1}+\re^{-\beta E_2}+\re^{-\beta E_5} 
\nonumber \\
\sum_m r(m) &\simeq& r(1)+r(2)+r(5) \nonumber \\
&\simeq& (H+\Delta_z)\re^{-\beta E_1}+
H\re^{-\beta E_2}+3H\re^{-\beta E_5}.
\nonumber \\
\eea
Hence $R_3(T)$ is approximated as 
\bea
R_3(T) &\simeq& 
\frac{\left(1+\frac{\Delta_z}{H}\right)\re^{-\beta E_1}+
\re^{-\beta E_2}+3\re^{-\beta E_5}}
{\tanh\left(\frac{\beta H}{2}\right)
\left(\re^{-\beta E_1}+\re^{-\beta E_2}+\re^{-\beta E_5}\right)} \nonumber \\
&\simeq& \frac{3+\re^{-\beta(H-H_{\rm c}+\Delta_z/2)}+
\left(1+\frac{\Delta_z}{H}\right)
\re^{-\beta(H-H_{\rm c}-\Delta_z/2)}}
{1+\re^{-\beta(H-H_{\rm c}+\Delta_z/2)}
+\re^{-\beta(H-H_{\rm c}-\Delta_z/2)}}.
\nonumber \\
\eea

In the absence of the DMI ($\Delta_z=0$), we have 
\be
R_3(T) \simeq 
\frac{3+2\re^{-\beta(H-H_{\rm c})}}
{1+2\re^{-\beta(H-H_{\rm c})}}.
\ee
Therefore, at ultra-cold temperatures 
($T\ll|H-H_{\rm c}|$), we have 
\be
R_3(T)\simeq\left\{\begin{array}{ll}
1+\re^{-|H-H_{\rm c}|/T} 
& (H<H_{\rm c}) \\
3-4\re^{-|H-H_{\rm c}|/T}
& (H_{\rm c}<H),
\end{array}\right.
\ee
and 
\be
\frac{\rd R_3(T)}{\rd T}\simeq\left\{\begin{array}{ll}
\frac{|H-H_{\rm c}|}{T^2}\,
\re^{-|H-H_{\rm c}|/T} 
& (H<H_{\rm c}) \\
-\frac{4|H-H_{\rm c}|}{T^2}\,
\re^{-|H-H_{\rm c}|/T} 
& (H_{\rm c}<H).
\end{array}\right.
\label{R3derivnoDM}
\ee

In the case of finite $\Delta_z$, at ultra-cold temperatures 
($T\ll\Delta_z$ for $H<H_{\rm c}-\Delta_z/2$ and 
$T\ll|H-H_{\rm c}-\Delta_z/2|$ for 
$H_{\rm c}-\Delta_z/2<H$), we have 
\begin{widetext}
\be
R_3(T)\simeq\left\{\begin{array}{ll}
1+\frac{\Delta_z}{H}-\frac{2\Delta_z}{H}\,
\re^{-\Delta_z/T} 
& (H<H_{\rm c}-\Delta_z/2) \\
1+\frac{\Delta_z}{H}+
\left(2-\frac{\Delta_z}{H}\right)\,
\re^{-|H-H_{\rm c}-\Delta_z/2|/T}
& (H_{\rm c}-\Delta_z/2<H<
H_{\rm c}+\Delta_z/2) \\
3-\left(2-\frac{\Delta_z}{H}\right)\,
\re^{-|H-H_{\rm c}-\Delta_z/2|/T} 
& (H_{\rm c}+\Delta_z/2<H),
\end{array}\right.
\ee
\end{widetext}
and 
\be
\frac{\rd R_3(T)}{\rd T}\simeq\left\{\begin{array}{ll}
-\frac{2\Delta_z^2}{T^2H}\,
\re^{-\Delta_z/T} 
& (H<H_{\rm c}-\Delta_z/2) \\
\frac{1}{T^2}\left(2-\frac{\Delta_z}{H}\right)
|H-H_{\rm c}-\frac{\Delta_z}{2}|\,
\re^{-|H-H_{\rm c}-\Delta_z/2|/T} 
& (H_{\rm c}-\Delta_z/2<H<
H_{\rm c}+\Delta_z/2) \\
\frac{-1}{T^2}\left(2-\frac{\Delta_z}{H}\right)
|H-H_{\rm c}-\frac{\Delta_z}{2}|\,
\re^{-|H-H_{\rm c}-\Delta_z/2|/T} 
& (H_{\rm c}+\Delta_z/2<H).
\end{array}\right.
\label{R3deriv}
\ee

In the uniaxial case, the structure of the energy levels is almost 
the same as that of ${\rm V}_{15}$ except that the gap at the 
avoided level crossing point closes.  
The equation explains the behavior of $R(T)$ in the simulation of 
${\rm V}_{15}$: 
$R'(T)<0$ at low fields, 
$R'(T)>0$ at fields close to the avoided level crossing, and 
$R'(T)<0$ at high fields.

\subsubsection{Weak Dzyaloshinsky-Moriya interaction}
\label{weakDMI}

Let us consider the full DMI case 
($D_x=D_y=D_z=D$).  We assume $D>0$ is small in the sense that 
$D \ll 2|H-H_{\rm c}|$ is satisfied.  

By ignoring smaller terms than $O(D)$, we have the following 
eigenvectors $\{\ket{u_j}\}$.
\bea
\ket{u_1} &=& -\sqrt{3}\re^{-\ri\pi/4}\xi\ket{v_5}+\ket{v_1} \nonumber \\
\ket{u_2} &=& \ket{v_5}+\sqrt{3}\re^{\ri\pi/4}\xi\ket{v_1} \nonumber \\
\ket{u_3} &=& -\re^{-\ri\pi/4}\eta\ket{v_2}+\ket{v_7} \nonumber \\
\ket{u_4} &=& -\sqrt{3}\re^{\ri\pi/4}\eta\ket{v_8}+\ket{v_3} \nonumber \\
\ket{u_5} &=& \ket{v_2}+\re^{\ri\pi/4}\eta\ket{v_7} \nonumber \\
\ket{u_6} &=& \ket{v_4}-\re^{-\ri\pi/4}\xi\ket{v_6} \nonumber \\
\ket{u_7} &=& \re^{\ri\pi/4}\xi\ket{v_4}+\ket{v_6} \nonumber \\
\ket{u_8} &=& \ket{v_8}+\sqrt{3}\re^{-\ri\pi/4}\eta\ket{v_3}, \nonumber \\
\eea
where 
\be
\xi\equiv\frac{-\Delta}{2\sqrt{2}(H-H_{\rm c})},\quad
\eta\equiv\frac{\Delta}{2\sqrt{2}(H+H_{\rm c})}.
\ee
The corresponding eigenvalues $\{E_j\}$ are given by 
\bea
E_8 &=& -\frac{3}{4}J+\frac{3}{2}H \nonumber \\
E_7 &=& -\frac{3}{4}J+\frac{1}{2}H \nonumber \\
E_6 &=& -\frac{3}{4}J-\frac{1}{2}H \nonumber \\
E_5 &=& -\frac{3}{4}J-\frac{3}{2}H \nonumber \\
E_4 &=& \frac{3}{4}J+\frac{1}{2}H+\frac{\Delta}{2} \nonumber \\
E_3 &=& \frac{3}{4}J+\frac{1}{2}H-\frac{\Delta}{2} \nonumber \\
E_2 &=& \frac{3}{4}J-\frac{1}{2}H+\frac{\Delta}{2} \nonumber \\
E_1 &=& \frac{3}{4}J-\frac{1}{2}H-\frac{\Delta}{2}.
\eea

From the above eigenvectors $\{\ket{u_j}\}$, 
nonzero matrix elements are obtained as
\bea
\bra{u_1}M^x\ket{u_4} &=& -\frac{1}{2} \nonumber \\
\bra{u_1}M^x\ket{u_6} &=& -2\re^{\ri\pi/4}\xi \nonumber \\
\bra{u_2}M^x\ket{u_3} &=& -\frac{1}{2} \nonumber \\
\bra{u_2}M^x\ket{u_4} &=& 0 \nonumber \\
\bra{u_2}M^x\ket{u_6} &=& \re^{-\ri\pi/4}\eta \nonumber \\
\bra{u_2}M^x\ket{u_8} &=& 0 \nonumber \\
\bra{u_3}M^x\ket{u_7} &=& -\eta\re^{-\ri\pi/4} \nonumber \\
\bra{u_4}M^x\ket{u_5} &=& -\sqrt{3}\re^{\ri\pi/4}\xi \nonumber \\
\bra{u_4}M^x\ket{u_7} &=& -\re^{\ri\pi/4}\xi \nonumber \\
\bra{u_5}M^x\ket{u_6} &=& \frac{\sqrt{3}}{2} \nonumber \\
\bra{u_6}M^x\ket{u_7} &=& 1 \nonumber \\
\bra{u_7}M^x\ket{u_8} &=& \frac{\sqrt{3}}{2}.
\eea

Let us consider the ultra-cold limit ($\beta\to\infty$), 
where all transitions can be ignored except for 
the transitions from the ground state.  
We consider two cases: 
$H< H_{\rm c}$, where 
$\ket{v_1}$ is the ground state, and 
$H_{\rm c}< H$, where 
$\ket{v_5}$ is the ground state.  We have
\bea
\sum_m\re^{-\beta E_m} &\simeq& 
\re^{-\beta E_1}+\re^{-\beta E_2} \nonumber \\
\sum_m r(m) &\simeq& r(1)+r(2) 
\simeq \left(H+\Delta\right)\re^{-\beta E_1} 
\left(H-\Delta\right)\re^{-\beta E_2} 
\nonumber \\
\eea
for $H<H_{\rm c}$ and 
\bea
\sum_m\re^{-\beta E_m} &\simeq& 
\re^{-\beta E_5}+\re^{-\beta E_1} \nonumber \\
\sum_m r(m) &\simeq& r(5)+r(1) \simeq 
3H\re^{-\beta E_5}+(H+\Delta)\,\re^{-\beta E_1}
\eea
for $H_{\rm c}<H$.  
Thus we obtain \eref{weakR3}.

% Create the reference section using BibTeX:
\bibliography{prbref}

\end{document}